\documentclass[twocolumn]{aastex62}
\usepackage{mathptmx}
\usepackage{bm}
\usepackage{amsmath}
\usepackage[utf8]{inputenc}
\usepackage{hyperref}
\usepackage{multirow}
\newcommand{\mathcircumflex}[0]{\mbox{\^{}}}

\shorttitle{Clusters in tidal fields}

\shortauthors{Meiron et al.}

\begin{document}

\title{Mass loss from massive globular clusters in tidal fields}

\correspondingauthor{Yohai Meiron}
\email{yohai.meiron@utoronto.ca}

\author[0000-0003-3518-5183]{Yohai Meiron}
\affiliation{Department of Astronomy and Astrophysics, University of Toronto, 50 St. George Street, Toronto, ON M5S\,3H4, Canada}
\affiliation{SciNet High Performance Computing Consortium, University of Toronto, 661 University Ave., Toronto, ON M5G\,1M1, Canada}

\author{Jeremy J. Webb}
\affiliation{Department of Astronomy and Astrophysics, University of Toronto, 50 St. George Street, Toronto, ON M5S\,3H4, Canada}

\author{Jongsuk Hong}
\affiliation{Korea Astronomy and Space Science Institute, 776 Daedeok-daero, Yuseong-gu, Daejeon 34055, Republic of Korea}

\author[0000-0003-4176-152X]{Peter Berczik}
\affiliation{National Astronomical Observatories and Key Laboratory of Computational Astrophysics,\\ Chinese Academy of Sciences, 20A Datun Rd., Chaoyang District, Beijing 100101, China}
\affiliation{Astronomisches Rechen-Institut, Zentrum für Astronomie der Universität Heidelberg, Mönchhofstr. 12--14, D-69120 Heidelberg, Germany}
\affiliation{Main Astronomical Observatory, National Academy of Science of Ukraine, 27 Akademika Zabolotnoho St., 03143 Kyiv, Ukraine}

\author{Rainer Spurzem}
\affiliation{Astronomisches Rechen-Institut, Zentrum für Astronomie der Universität Heidelberg, Mönchhofstr. 12--14, D-69120 Heidelberg, Germany}
\affiliation{National Astronomical Observatories and Key Laboratory of Computational Astrophysics,\\ Chinese Academy of Sciences, 20A Datun Rd., Chaoyang District, Beijing 100101, China}
\affiliation{Kavli Institute for Astronomy and Astrophysics, Peking University, Yiheyuan Lu 5, Haidian District, Beijing 100871, China}

\author[0000-0002-7667-0081]{Raymond G. Carlberg}
\affiliation{Department of Astronomy and Astrophysics, University of Toronto, 50 St. George Street, Toronto, ON M5S\,3H4, Canada}

\begin{abstract}
Massive globular clusters lose stars via internal and external processes. Internal processes include mainly two-body relaxation, while external processes include interactions with the Galactic tidal field. We perform a suite of $N$-body simulations of such massive clusters using three different direct-summation $N$-body codes, exploring different Galactic orbits and particle numbers. By inspecting the rate at which a star's energy changes as it becomes energetically unbound from the cluster, we can neatly identify two populations we call kicks and sweeps, that escape through two-body encounters internal to the cluster and the external tidal field, respectively. We find that for a typical halo globular cluster on a moderately eccentric orbit, sweeps are far more common than kicks but the total mass loss rate is so low that these clusters can survive for tens of Hubble times. The different $N$-body codes give largely consistent results, but we find that numerical artifacts may arise in relation to the time step parameter of the Hermite integration scheme, namely that the value required for convergent results is sensitive to the number of particles.
\end{abstract}


\section{Introduction}

Globular clusters are dynamical systems with complex evolutionary histories. Milky Way globular clusters are estimated to have formed between 13 and 9 Gyr ago \citep{VandenBerg+2013}. This globular cluster population is thought to be a mixture of objects that have formed \textit{in situ} and objects that have been accreted as the Galaxy grew (cf. \citealt{Cote+1998} on globular cluster systems in giant elliptical Galaxies). Their subsequent evolution is then strongly linked to their mass.

Early in a cluster's lifetime, mass is primarily lost through stellar evolution, where gas is expelled from the cluster as its member stars evolve. In addition to stellar evolution, individual stars may escape from the cluster through stellar dynamical processes, thereby reducing its total mass. These processes can be broadly categorized as internal and external. (Internal) two-body encounters \citep{Chandrasekhar1942} may result in stars gaining enough speed to escape the cluster, while the (external) Galactic tidal field leads to a truncation surface for the gravitational influence of the cluster \citep{vonHoerner1957}, and stars that find themselves outside of it have effectively escaped as well. These mechanisms are coupled as we show in this work. An additional and potentially important internal mass loss mechanism is triple (binary--single) interactions; it is not considered in this work.

In many globular clusters, all these processes are coupled because the timescales associated with them are comparable. This situation is different from dwarf galaxies, for example, where the two-body timescale far exceeds the tidal timescale, or in open clusters where the opposite is true. Due to the coupling of these two time scales, the problem of globular cluster mass loss, and hence their survivability, is more difficult to model; thus numerical simulations are the primary tool.

\citet{Chernoff+1990} have used Fokker-Planck modelling (following from \citealt{Rosenbluth+1957} and \citealt{Henon1961}) to numerically study the evolution of a globular cluster on a circular orbit, approximating the Galaxy as a point mass. This was followed by simplified and full $N$-body simulations \citep{Fukushige+1995, Vesperini+1997, Portegies-Zwart+1998}, Monte Carlo models \citep{Joshi+2001, Giersz2001}, and more sophisticated (anisotropic) Fokker-Planck models \citep{Takahashi+1998, Takahashi+2000, Takahashi+2000b}.

A circular orbit around a point mass Galaxy results in a constant tidal field, which is relatively simple to model, but does not reflect the reality of Galactic globular cluster orbits \citep[e.g.][]{Frenk+1980, Dinescu+1999, Vasiliev2019}. An early numerical study of globular clusters on eccentric orbits was performed by \citet{Oh+1992}. Their hybrid numerical method accounted for both two-body relaxation (through orbit-averaged diffusion coefficients) and the time-varying Galactic tides (through direct integration). They concluded that while the limiting radius does not depend on the orbital phase, it is strongly influenced by two-body relaxation: being close to the value of the Jacobi radius at pericentre if the relaxation timescale is long, or close to the apocentre value if the relaxation time is short; these results, however, have been contested in more recent studies (see below).

\citet{Baumgardt+2003} further explored the effects of eccentricity by performing a large set of $N$-body simulations. Their globular cluster models varied by particle numbers and density profiles, and were put on eccentric orbits around a singular isothermal sphere (rather than a point mass) representing the Galaxy. They accounted for stellar evolution and showed that its effect is to almost instantaneously decrease the cluster mass by 30 per cent. They also showed that the dissolution time of a cluster on an eccentric orbit has a simple relation to the dissolution time of a cluster on a circular orbit. \citet{Cai+2016} write this relation as
\begin{equation}
T_\mathrm{diss}(a,\epsilon)=(1-\epsilon^2)(1-c\epsilon^2)T_\mathrm{diss}(a,0)\label{eq:Cai}
\end{equation}
where $a$ is the semi-major axis, $\epsilon$ is the eccentricity and $c$ is a constant. For \citet{Baumgardt+2003}, $c = 0$ provides a good fit, but \citet{Cai+2016} find $c \sim 0.5$ for their simulations, which differed in several aspects, including the Galaxy model.

In the past decade more information has been gathered about the effects of an eccentric orbit on a cluster's structure and size. \citet{Kupper+2010} found that the limiting radius of a cluster adjusts to the mean Jacobi radius along the orbit, while \citealt{Webb+2013} further showed that a cluster does not need to fully relax in order to expand, and that the limiting radius nearly traces the instantaneous tidal radius of the current orbital phase (which is more apparent for high eccentricities).

Past studies have focused on modest-sized clusters that could be simulated fiducially (i.e. with one particle representing one star) using $N \sim 10^5$. Additionally, each study has used a single $N$-body code to produce its results. Here, we present a series of simulations of relatively massive ($\sim 5\times 10^5\,\mathrm{M}_\odot$) globular clusters, which we simulate with up to $N = 10^6$ particles. We use multiple $N$-body codes to study the cluster's evolution as well as the codes' characteristics and the effect of softening. We test circular and moderately eccentric orbits and two representative values for the semi-major axis for weak and strong tides.

In Section \ref{sec:methods} we explain the initial conditions and codes used and present the cluster models. In Section \ref{sec:weak-unsoftened} we discuss the results from several of our models that were not tidally dominated (but rather relaxation dominated), we call this the ``weak tides'' case. In Section \ref{sec:strong} we discuss the opposite case where the tidal field dominates the mass evolution. In Section \ref{sec:fit} we attempt to use a simple formula for the mass loss as a template to fit the mass evolution.

\section{Methods}\label{sec:methods}

\subsection{Initial conditions}

Our clusters' density profile follows a \citet{King1966} model with $W_0=5$ and a half-mass radius of $r_\mathrm{h}=19\,\mathrm{pc}$ (and thus truncation or `tidal' radius of $r_\mathrm{t}=102\,\mathrm{pc}$) produced with the \textsc{mcluster} code \citep{Kupper+2011}. We consider models with $N=10^5$ and $10^6$ stars. The stellar masses are produced by evolving a \citet{Kroupa2001} initial mass function with initial masses between 0.1 and $50\,\mathrm{M}_\odot$ for 300 Myr (the matallicity is $Z=10^{-4}$); for the $N=10^6$ models this evolution corresponds to masses that range between 0.1 and $3.1\,\mathrm{M}_\odot$. The total mass is $4.39\times 10^5\,\mathrm{M}_\odot$ (for the $N=10^5$ models, each mass is multiplied by 10, so the total cluster mass is the same). There are no primordial binaries or binary evolution. There is no further stellar evolution after the initial setup. The cluster is then initialized at the apocentre position of its orbit at $t=0$. Two Galactic orbits are explored in addition to an isolated cluster (Section~\ref{sec:simulations}).

The Galactic potential is smooth (lacks substructure) and spherically symmetric, given by
\begin{equation}
\Phi_{\mathrm{ext}}(r)=v_{0}^{2}\log(r/r_{0})
\end{equation}
where $v_{0}=240\,\mathrm{km\,s^{-1}}$ and $r_{0}=1\,\mathrm{kpc}$. Dynamical friction is not considered in this work, so that the centre of density of each cluster has a periodic orbit.

\subsection{Codes}\label{sec:codes}

We use three codes to simulate the cluster's evolution in a tidal field: \textsc{nbody6++} \citep{Wang+2015}, $\varphi$\textsc{grape} (\citealt{Harfst+2007}, Meiron et al., in preparation), and \textsc{\textsc{ph4}} \citep{McMillan+2012}. These codes are independent implementations but have many similarities. All three codes are parallel, GPU-accelerated direct $N$-body codes, they are all fourth-order Hermite integrators \citep{Makino1991} that employ block time steps based on the Aarseth time step criterion (see e.g. \citealt{Aarseth2003}):
\begin{equation}
\Delta t=2\mathcircumflex\left\lfloor \frac{1}{2}\log_{2}\left(\eta\frac{|\bm{a}||\ddot{\bm{a}}|+|\dot{\bm{a}}|^{2}}{|\dot{\bm{a}}||\dddot{\bm{a}}|+|\ddot{\bm{a}}|^{2}}\right)\right\rfloor \label{eq:time-step}
\end{equation}
where $\bm{a}$ is the star's acceleration and the dots represent derivatives with respect to time; $\left\lfloor \cdots\right\rfloor$ is the rounding down operator that is needed so that the time steps occur in discrete blocks rather than as a continuum (additionally, the step size in this scheme can only increase or decrease by a factor of two, and must also be commensurate with the current integration time; these requirements are needed to facilitate parallel computing); and $\eta$ is the time step parameter.

$\varphi$\textsc{grape} and \textsc{ph4} are quite simple integrators that optionally use force softening to prevent the formation of hard binaries. Integration of hard binaries is very costly, and a spontaneous formation of such a pair can slow the simulation to a halt. Other than the implementation, these two codes differ in how the external (Galactic) force is applied: in $\varphi$\textsc{grape} the external force is calculated for every active particle at every time step, while \textsc{ph4} is used within the AMUSE framework \citep{Pelupessy+2013,Portegies_Zwart+2018} and the external force is applied through the Bridge scheme \citep{Fujii+2007}, where each star's velocity is changed at constant intervals that are much longer than the integration time step. \textsc{nbody6++} does not employ force softening, but instead uses the \citet{Kustaanheimo+1965} regularization technique to treat binaries. Additionally, it uses the \citet{Ahmad+1973} neighbour scheme to reduce the number of full force calculations required. The external force is applied in much the same way as in $\varphi$\textsc{grape}: at every time step, the active stars are temporarily moved from the centre-of-mass frame of reference to the Galactic frame of reference where the force is calculated (in some other versions of this code, the centrifugal and Coriolis forces are calculated in the cluster's frame of reference; this is suitable for circular orbits only, unless the Euler force is included as well).

\subsection{Simulations}\label{sec:simulations}

\begin{table}
\centering
\begin{tabular}{|l|c|c|c|c|l|}
\hline  Model & $a$ & $N$ & $e$ & soft & code \\
\hline \hline  \texttt{1} & 19.7 & $10^6$ & 0.52 & no & \textsc{nbody6++}\\
\hline  \texttt{6} & 19.7 & $10^6$ & 0.52 & yes & $\varphi$\textsc{grape}\\
\hline  \texttt{17} & 19.7 & $10^5$ & 0.52 & no & \textsc{nbody6++}\\
\hline  \texttt{18} & 19.7 & $10^5$ & 0.52 & no & $\varphi$\textsc{grape}\\
\hline  \texttt{19} & 19.7 & $10^5$ & 0.52 & no & \textsc{ph4}\\
\hline  \texttt{22} & 19.7 & $10^5$ & 0.52 & yes & $\varphi$\textsc{grape}\\
\hline  \texttt{25} & 19.7 & $10^5$ & 0 & no & \textsc{nbody6++}\\
\hline  \texttt{33} & 3.35 & $10^6$ & 0.49 & no & \textsc{nbody6++}\\
\hline  \texttt{38} & 3.35 & $10^6$ & 0.49 & yes & $\varphi$\textsc{grape}\\
\hline  \texttt{49} & 3.35 & $10^5$ & 0.49 & no & \textsc{nbody6++}\\
\hline  \texttt{54} & 3.35 & $10^5$ & 0.49 & yes & $\varphi$\textsc{grape}\\
\hline  \texttt{57} & 3.35 & $10^5$ & 0 & no & \textsc{nbody6++}\\
\hline  \texttt{iso} & --- & $10^5$ & --- & no & \textsc{nbody6++}\\
\hline
\end{tabular}
\caption{List of models. $a$ is given in kpc.\label{tab:models}}
\end{table}

We present thirteen models that differ in the Galactic orbit, the number of particles $N$, softening, and code used. The models are summarized in Table~\ref{tab:models}; some of the models have multiple simulations with different random seeds (not shown in the table). The first seven models in the Table are on a Galactic orbit with semi-major axis of 19.7~kpc. This orbit leads to a relatively weak tidal field. This tidal field is considered weak because for every point along the orbit, the initial King model's truncation radius is smaller than the Jacobi radius, which we na\"ively define as
\begin{equation}
r_\mathrm{J} = R \left( \frac{M}{3M_\mathrm{enc}} \right)^{1/3}\label{eq:tidal-radius}
\end{equation}
where $R$ is the distance from the centre of the Galactic model and $M_\mathrm{enc}$ is the Galactic mass enclosed within $R$. The following five models in Table~\ref{tab:models} differ from the above by having a much smaller Galactic orbit with semi-major axis of 3.35~kpc, leading to a strong tidal field.

For the non-circular models in the weak and strong tidal cases, the starting points of the clusters are at the apocentres of 30 and 5~kpc, respectively, and the initial velocities are $120~\mathrm{km\,s}^{-1}$ in the positive $y$ direction in both cases. The eccentricities in these orbits are 0.52 and 0.49, respectively. The `soft' column indicates whether gravitational softening was applied; if yes, the softening length was 0.0024~pc. Finally, the model in the last row is in isolation (i.e. not in any tidal field). Note that the numbers assigned to the models do not indicate the chronological order or that some models are not shown, rather they encode some of the information about each model to ease bookkeeping.

\section{Weak tides case}\label{sec:weak-unsoftened}

\subsection{Unsoftened models}\label{sec:reference-models}

As our reference models for the evolution in a weak tidal field, we take the unsoftened models using $10^5$ and $10^6$ stars evolved with \textsc{nbody6++} (models \texttt{1} and \texttt{17}, respectively). Later in this section we compare model \texttt{17} to models \texttt{18} and \texttt{19} that are performed with the other codes.

\begin{figure}
\centering \includegraphics[width=1\columnwidth]{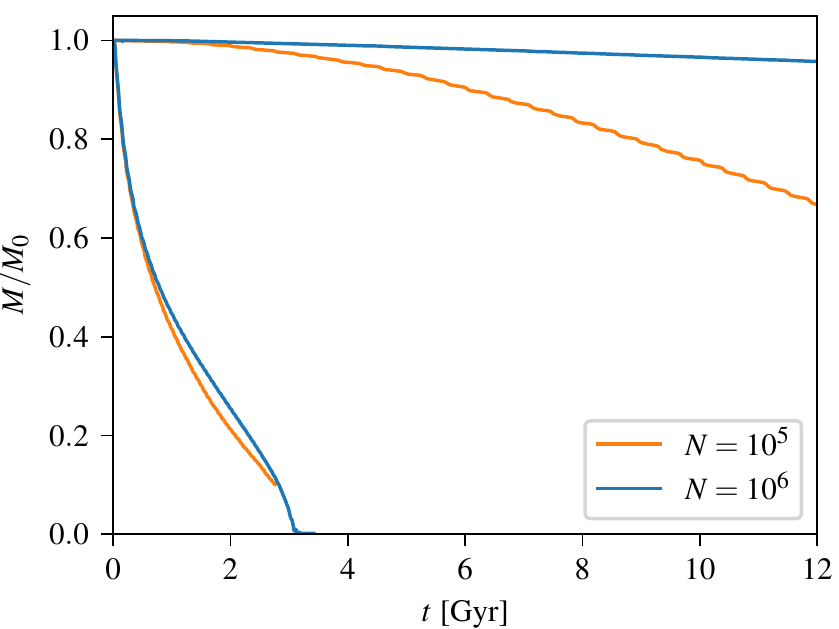}
\caption{Remaining bound mass of the cluster as a function of time for models in an orbit experiencing a weak tidal field (upper two curves) and an orbit experiencing a strong tidal field (lower two curves). The blue and orange curves represent models with $N=10^5$ and $10^6$ stars, respectively.}
\label{fig:mass-loss}
\end{figure}

Figure~\ref{fig:mass-loss} shows the remaining bound mass fraction as a function of time. A bound star is defined as a star that has negative \textit{internal} energy. The internal energy is the sum of the potential energy due to all other stars and kinetic energy with respect to the centre of mass, per unit mass.\footnote{The definition of a bound star as a star with negative internal energy is only strictly correct in an isolated cluster. On a circular Galactic orbit, the potential at the Lagrangian points determines the escape energy threshold unambiguously. On an eccentric Galactic orbit, on the other hand, the potential at the effective potential's saddle points changes substantially over an orbital period, and especially quickly around the pericentre passage. Stars often find themselves with internal energies higher than the potential at a saddle point but only for a brief time. The choice of zero as the escape threshold is therefore a practical one.} The centre of mass itself should be calculated from bound stars only, thus an iterative process is needed. Since the internal energy does not include a term for the Galactic potential, it is not a conserved quantity for the cluster as a whole. The top blue and orange curves show the remaining bound mass for models \texttt{1} and \texttt{17}, respectively. Model \texttt{1} lost 4 per cent of its mass over 12 Gyr, while model \texttt{17} lost 33 per cent over that time. The top blue and orange curves show the remaining bound mass for models \texttt{33} and \texttt{49}, respectively, which are the equivalents of the formerly discussed pair of models, but in the strong-tide case discussed in Section \ref{sec:strong}.

\begin{figure}
\centering \includegraphics[width=1\columnwidth]{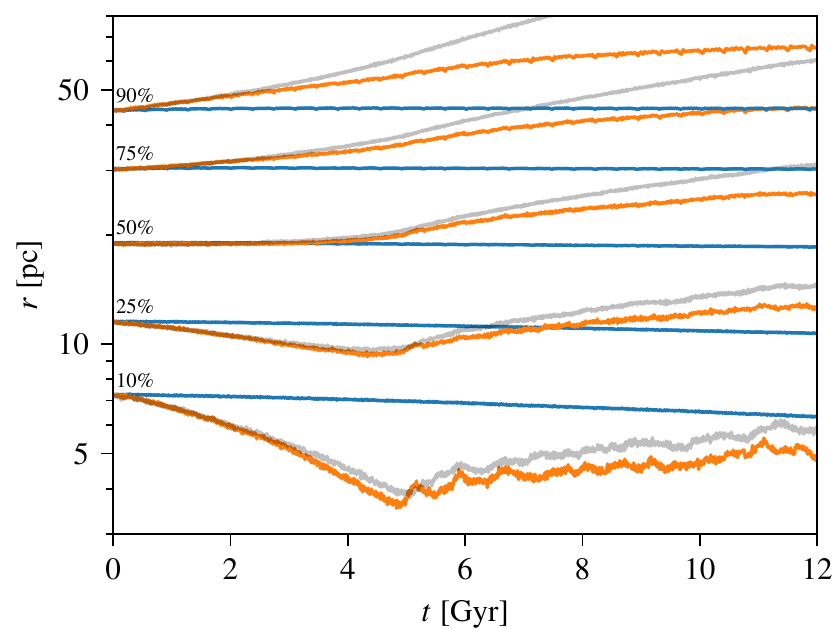}
\caption{Selected Lagrangian radii for models on an eccentric orbit in a weak tidal field, model \texttt{1} (blue) with $N=10^6$ and \texttt{17} (orange) with $N=10^5$. For reference the isolated model \texttt{iso} is also shown. The central regions of the cluster are `protected' from the tidal field and the evolution is dominated by two-body encounters alone.}
\label{fig:lagrange-radii-weaktide}
\end{figure}

To understand mass loss it is also important to look at the internal structure of the cluster. Figure~\ref{fig:lagrange-radii-weaktide} shows five representative Lagrange radii of the two models in blue and orange. These show the radii that enclose 10, 25, 50, 75, and 90 per cent of the remaining bound mass at each time. While these radii stay nearly constant for model \texttt{1} with $N=10^6$, model \texttt{17} with $N=10^5$ shows a more significant evolution, including core collapse after about 5 Gyr. The 10 per cent Lagrange radius decreases by a factor smaller than 0.15 throughout the former simulation, however its characteristic shape is very similar to that of the $N=10^5$ model. That could be seen by ``slowing down'' time for model \texttt{17}; we find numerically that the best fit ($R^2=0.98$) is obtained when the time is scaled by a factor of 8.2. Similarly we can try to match the 25 per cent Lagrange radius, the best fit ($R^2=0.95$) is obtained for a factor of 7.4. This approach is more difficult for the larger Lagrange radii, as the behaviour is qualitatively different.

The best fitting scaling factor for the 10 per cent Lagrange radius could be understood from the scaling of the two-body relaxation timescale \citep{Spitzer1987} with the number of particles, namely $t_\mathrm{relax} \sim N/\ln(0.4 N)$, this gives a ratio of 8.2 between the two models. The fact that this only works at the very inner regions of the cluster tells us that the core is relatively protected from influence of the tidal field, and evolve in the same way as if the clusters were in isolation. This point is further strengthened by inspecting the behaviour of the isolated model \texttt{iso} (that has the same number of particles as model \texttt{17}), shown in grey in Figure~\ref{fig:lagrange-radii-weaktide}. The isolated cluster becomes significantly larger in the outer Lagrange radii because there is no Jacobi surface, and stars can still be bound even if they are extremely distant from the cluster's centre.

\begin{figure}
\centering \includegraphics[width=1\columnwidth]{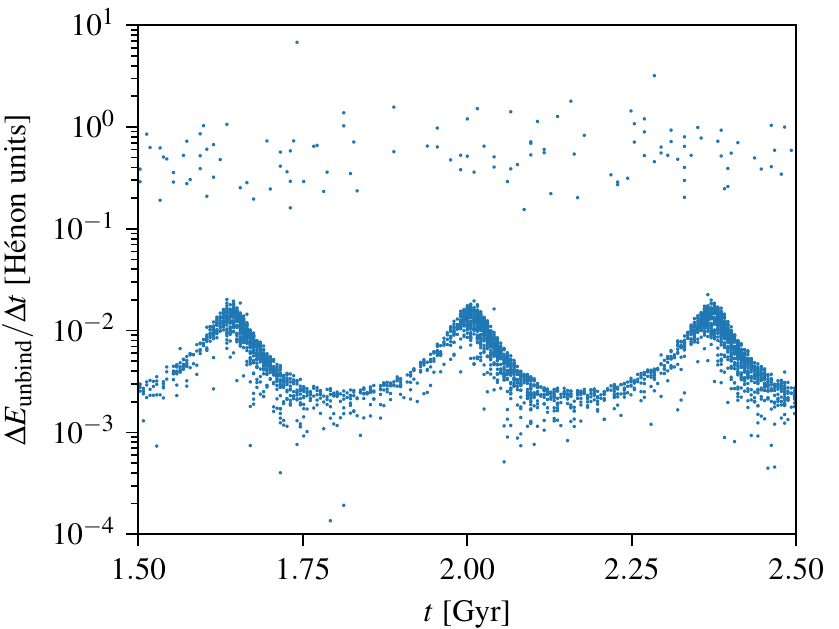}
\caption{Each points in the scatter plot represents a star that escaped the cluster in model \texttt{1} that has $N=10^6$ particles and represent the weak tidal field case. The time $t$ is of each star's escape, and the quantity on the vertical axis is the difference in internal energy of the star between the snapshots immediately before and after it has become unbound, divided by the time difference between the two snapshots (which is a constant in this simulation). The two distinct populations represent \textit{kicks} (stars that escape due to a two-body encounter) and \textit{sweeps} (stars that escape due to the tidal forces), at high and low values of $\Delta E_\mathrm{unbind}$, respectively. Figure~\ref{fig:kick-sweep-example} shows an example of an orbit from each population.}
\label{fig:kick-sweep-statistics}
\end{figure}

\begin{figure}
\centering \includegraphics[width=1\columnwidth]{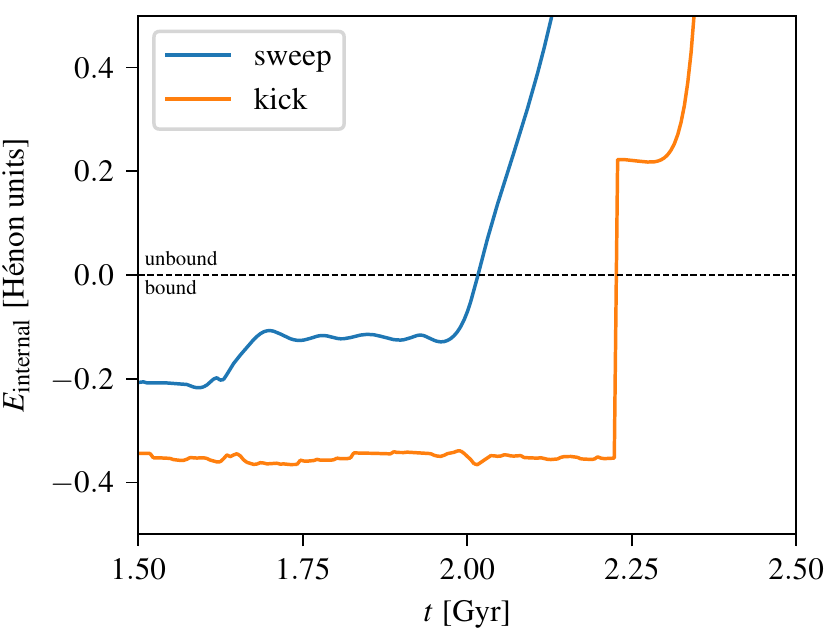}
\caption{The internal energy as a function of time for representative examples of a sweep (blue) and a kick (orange). The former's internal energy increases over tens of millions of years due to the tidal forces, while the latter receives a very large and quick boost to its velocity (and thus kinetic energy) due to a two-body encounter. The quantity plotted in Figure~\ref{fig:kick-sweep-statistics} is the slope at the moment the internal energy crosses zero. Notice that kicks generally come from deeper in the potential well than sweeps (where the average time between encounters is shorter).}
\label{fig:kick-sweep-example}
\end{figure}

While Figure~\ref{fig:lagrange-radii-weaktide} explores the structure of the remaining bound mass, Figure~\ref{fig:kick-sweep-statistics} focuses on escaping stars. It is a scatter plot of all stars in model \texttt{1}, that have escaped (i.e. have become unbound) between 1.5 and 2.5 Gyr. The quantity on the vertical axis is the difference in internal energy of the star between the snapshots immediately before and after it has become unbound, $\Delta E_\mathrm{unbind}$, divided by the time difference between the two snapshots, $\Delta t$ (which is a constant in this simulation). This quantity is in H\'enon units (it is not particularly useful to convert it to physical units), it corresponds to the slope of the energy curve at escape time (see representative examples in Figure~\ref{fig:kick-sweep-example}). Only a short representative time interval of 1 Gyr is shown in order not to crowd the plot. For this model, the pattern repeats itself and the point density remains fairly constant.

Two distinct populations are immediately apparent in this scatter plot, distinguished by their $\Delta E_\mathrm{unbind}$ values. We call the population with higher and lower values \textit{kicks} and \textit{sweeps}, respectively. Kicks are stars that got a sufficiently large velocity boost to escape the cluster due to a two-body encounter. Sweeps are stars that have received more gradual velocity increment due to the tidal forces until their internal energies become positive. The sweep rate is quasiperiodic with the Galactic orbit, the orbitally averaged rate is nearly constant for model \texttt{1} and decreasing for model \texttt{17}. Kicks are relatively rare, accounting for 2.69 per cent of escapes in model \texttt{1} over the simulation time of 12 Gyr.

To illustrate the difference between kicks and sweeps, Figure~\ref{fig:kick-sweep-example} shows the internal energies of two stars representative of these two populations as a function of time. The reason for the separation of the two populations is immediately apparent in this plot: the energy to escape the cluster is gained on two very different time scales in the two cases. In the case of a sweep, the internal energy increases over tens of millions of years prior to the star becoming unbound, at roughly a constant rate (a pericentre passage of the cluster around the Galaxy occurs at 2.008~Gyr). This is the same order of magnitude as the star's orbital period, as prior to its escape, it hovers at around $\sim 100$~pc from the centre of the cluster ($\sim 5 \times$ the half-mass radius). In the case of a kick, the internal energy increases very quickly, so much that the energy curve appears discontinuous (the snapshots are taken 5~Myr apart). The relevant timescale in this case is that of the two-body encounter, that can be as short as a few years.

\begin{figure}
\centering \includegraphics[width=1\columnwidth]{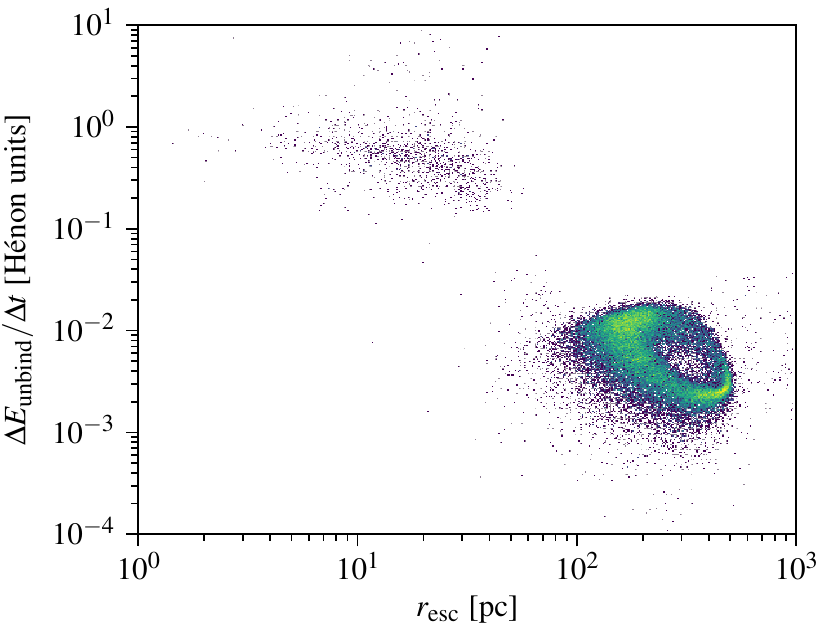}
\caption{The $y$-axis shows the same quantity as in Figure~\ref{fig:kick-sweep-statistics} for all escapers in model \texttt{1} (regardless of time of escape), while the $x$-axis shows the last known bound radius within the cluster, i.e. the radius in the snapshot just prior to the star becoming unbound. Green and yellow hues indicate a higher density of points (in order to avoid saturation), but the colour carries no meaning otherwise. Kicks and sweep can be distinguished by radius as well.}
\label{fig:kick-sweep-radius}
\end{figure}

In Figure~\ref{fig:kick-sweep-example}, the kick came from deeper in the potential well than the sweep, and this is generally the case because the denser inner regions have a higher rate of two-body encounters. Figure~\ref{fig:kick-sweep-radius} illustrates this point better by showing $\Delta E_\mathrm{unbind}/\Delta t$ as a function of the last known bound radius within the cluster, i.e. the radius in the snapshot just prior to the star becoming unbound. The plot includes all escapers in model \texttt{1} (regardless of time of escape). We see that one can even distinguish kicks and sweeps solely based on the last bound radius before escape, although there is some small overlap. The transition is at $\sim40\,$pc, which is $\sim$ twice the initial half-mass radius. Initially, just 13 per cent of the mass lies beyond that radius. In the absence of any relaxation, this would be the reservoir for sweeps.

Model \texttt{17}, that has ten times fewer particles, has a fairly similar kick/sweep behaviour, with kicks accounting for 2.68 per cent of all escapes. The similarity of this number to the model \texttt{1} kick fraction is somewhat coincidental. If, for example, only escapes that occur before 10~Gyr are considered (instead of 12~Gyr), the kick fractions are 2.8 and 3.3 per cent for models \texttt{1} and \texttt{17}, respectively. As noted above, the kick distribution in model \texttt{1} is fairly constant in time. Model \texttt{17} however has a varying kick rate. Additionally, the $\Delta E_\mathrm{unbind}$ distribution changes post-core collapse, gaining a heavy high-energy tail.

\begin{figure}
\centering \includegraphics[width=1\columnwidth]{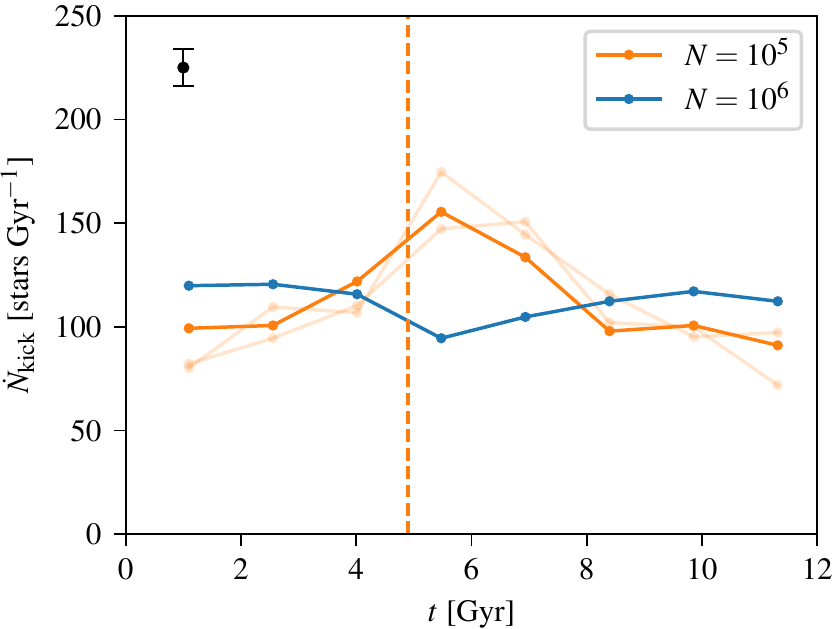}
\caption{Kick rates for models \texttt{1} ($N=10^6$; blue) and \texttt{17} ($N=10^5$; orange; two additional simulations with identical parameters but different random seeds are shown in light orange). The dashed vertical orange line corresponds to the core collapse time for model \texttt{17}. While the blue curve remains constant to within 30 per cent, orange curves change by up to a factor of 2. The black symbol at top left shows a typical errorbar based on the square root of the number count.}
\label{fig:kick-rate-time}
\end{figure}

Figure~\ref{fig:kick-rate-time} shows the rate of kicks as a function of time for the two models in blue and orange. Due to the overall small number of kicks, the rate has to be calculated over wide time bins. In order to show the statistical significance, we also plot the kick rate for two additional simulations (with different random seeds) of model \texttt{17} in light orange. We only have one simulation for model \texttt{1} due to its very high computational cost, however it is likely that the dip around 6~Gyr is a random fluctuation, as nothing special happens to this model at this time.

While the kick rates for both models are comparable in order of magnitude, we note that the particles in model \texttt{17} are 10 times more massive on average, leading to larger overall mass lost from kicks. In physical units, over 12~Gyr model~\texttt{1} loses 530 and 19\,000~$\mathrm{M}_\odot$ from kicks and sweeps, respectively, while the corresponding numbers for model~\texttt{17} are 4\,300 and 155\,000~$\mathrm{M}_\odot$.

\subsection{Code comparison}

\begin{figure}
\centering \includegraphics[width=1\columnwidth]{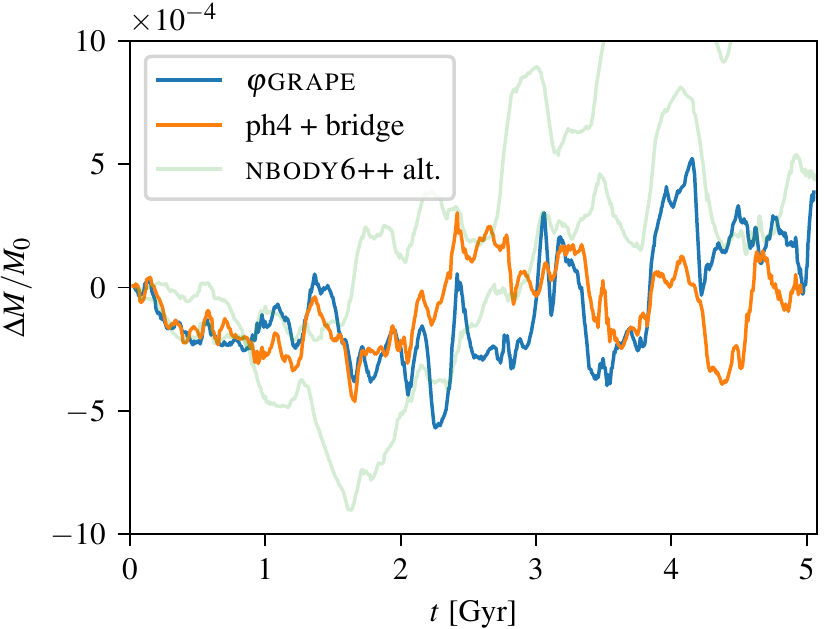}
\caption{Comparison of models performed with different $N$-body codes. The blue and orange lines are the differences in the remaining bound masses between models \texttt{18} and \texttt{19} on one hand (performed with $\varphi$\textsc{grape} and \textsc{ph4}, respectively) and \texttt{17} on the other (performed with \textsc{nbody6++}). The light green lines are similarly the differences between models \texttt{17A} and \texttt{17B} on one hand and \texttt{17} on the other, that only differ in random seed. These models are in our weak tidal field eccentric orbit and have $N=10^5$ stars. No gravitational softening was used. The small and seemingly random residuals indicate that there is no systematic difference in these models.}
\label{fig:comparison-codes}
\end{figure}

The previous section presented models \texttt{1} and \texttt{17}, which were produced with \textsc{nbody6++} and differed by the number of stars. Model \texttt{17} with $N=10^5$ stars was repeated with two other codes: models \texttt{18} and \texttt{19} that were performed with $\varphi$\textsc{grape} and \textsc{ph4}, respectively. The results were practically identical. These simulations were shorter and ended at $\sim 5$~Gyr. As noted in Section~\ref{sec:codes}, these two codes can stall when used without softening due to the formation of hard binaries. In model \texttt{18} we had to manually separate two hard binaries over the course of the simulation while in model \texttt{19} we had to separate one over this time period. Figure~\ref{fig:comparison-codes} shows the difference in remaining bound mass between models \texttt{18} and \texttt{19} on one hand, and \texttt{17} on the other. The relative difference does not exceeded $5\times 10^{-4}$, and seems to be random in nature, indicating that there is no systematic difference. The light green lines are similarly the differences between models \texttt{17A} and \texttt{17B} on one hand and \texttt{17} on the other that only differ in random seed. Apparently the difference in realization may cause a bigger difference in the mass evolution (however, still very small) than using the same realization but with a different code.

Despite the differences between \textsc{nbody6++}, $\varphi$\textsc{grape}, and \textsc{ph4}, we find that the evolution of $N=10^5$ clusters in weak tidal fields is nearly identical from one code to the next.

\subsection{Softened models}\label{sec:weak-softened}

Gravitational softening is a technique to ease the computation of $N$-body simulation. In ``Plummer''-type softening, the gravitational force between two bodies is modified from $\sim r^{-2}$ to $\sim (r^2 + a^2)^{-1}$, where $a$ is called the softening length. For $r\gg a$ the modified force approaches the true Newtonian force, but otherwise it approaches a constant. This hinders the formation of bound systems on scales $\leq a$ which otherwise have small timesteps and are difficult to integrate unless a regularization technique is used. Modifying the gravitational force in such a way has implications for hyperbolic encounters as well, the effect is decreasing the Coulomb logarithm and thus increasing of the two-body relaxation time.

\subsubsection{Model with $N=10^5$}

Model \texttt{22} is the softened equivalent of model \texttt{17} presented in Section~\ref{sec:reference-models}. The two models evolve in a very similar way, and most quantities (total mass loss, Lagrange radii) match when \texttt{17} is ``slowed down'' by a factor of 1.18. Between 0.2 and 5~Gyr (roughly the time of the core collapse), model \texttt{17} experiences 461 kicks and 8\,593 sweeps, while model \texttt{22} experiences 129 kicks and 6\,620 sweeps. The softer potential causes a significant reduction in the number of kicks (as expected given the treatment of close encounters) and a moderate reduction in the number of sweeps.

\subsubsection{Model with $N=10^6$}\label{sec:bad-model}

\begin{figure}
\centering \includegraphics[width=1\columnwidth]{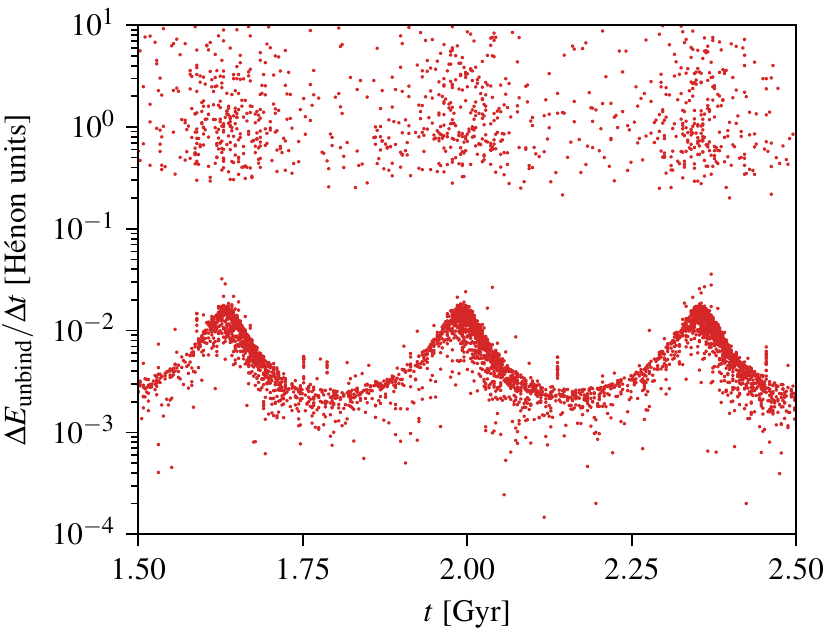}
\caption{Same as Figure~\ref{fig:kick-sweep-statistics} but for model \texttt{6} which is has the same initial conditions as model \texttt{1} used in that Figure, but used the $\varphi$\textsc{grape} code with gravitational softening length of 0.0024~pc. The main difference is that here the kick rate is not constant along one orbital period, but higher around pericentre.}
\label{fig:kicks-sweeps-statistics-fail}
\end{figure}

Model \texttt{6} is the softened equivalent of model \texttt{1} presented in Section~\ref{sec:reference-models}. It behaves unexpectedly. While it is similar to model \texttt{1}, one would expect a somewhat lower mass loss rate due to the gravitational softening, as seen when comparing models \texttt{22} and \texttt{17}. However the result is the opposite. The difference in kick rate is especially stark: Between 0.2 and 5~Gyr, model \texttt{1} experiences 558 kicks and 14\,530 sweeps, while model \texttt{6} experiences 5\,369 kicks and 16\,607 sweeps. This $\sim$ tenfold increase in the kick rate (as opposed to a decrease by a factor of 3.6 when softening is added in the $N=10^5$ case) seem to indicate numerical error.

While the kick rate in model \texttt{1} is uniform in time, in model \texttt{6} the kicks concentrate around pericentre as seen in Figure~\ref{fig:kicks-sweeps-statistics-fail}. This signature suggests that numerical error in the integration process is the cause of the excess in kicks in this model, as there should be no dependence between kick rate and orbital phase. We speculate that the simulation has not converged with respect to the time step parameter $\eta$ (see Equation~\ref{eq:time-step}). The small vertical clusters of points seen at e.g. $t=1.75$, $2.10$, and $2.45\,\mathrm{Gyr}$ are an artifact due to several missing snapshots.

Convergence has been tested with $\varphi$\textsc{grape} for $N=10^5$ and it was found that models with $\eta=0.01$ and $0.005$ were indistinguishable, while a model with $\eta=0.02$ lost mass at measurably higher rate (its dissolution time was $\sim 15$ per cent shorter). It was therefore assumed that the integration is accurate enough for $\eta=0.01$ and that this does not depend on $N$. Since convergence tests with $N=10^6$ are extremely computationally expensive, the 0.01 value was used for model \texttt{6}.

However, the pattern seen in Figure~\ref{fig:kicks-sweeps-statistics-fail} (specifically the increase in kick rate around pericentre) is unexpected. Other than in model \texttt{1} it is only seen in our version of model \texttt{17} that is performed with $\eta=0.02$ mentioned above, that we established did not converge with respect to $\eta$. We note that the \textsc{nbody6++} models were all performed with $\eta=0.02$, however due to the neighbour scheme, this code's behaviour with respect to $\eta$ may be significantly different to $\varphi$\textsc{grape}'s.

It is unclear exactly how the orbital phase-dependant excess in kicks is produced, and why the value of $\eta$ required for a converging result depends on $N$. The softening length and time step parameter may also interact in non-trivial ways that only become apparent in clusters with $N \sim 10^6$ stars.

\section{Strong tides case}\label{sec:strong}

\begin{figure}
\centering \includegraphics[width=1\columnwidth]{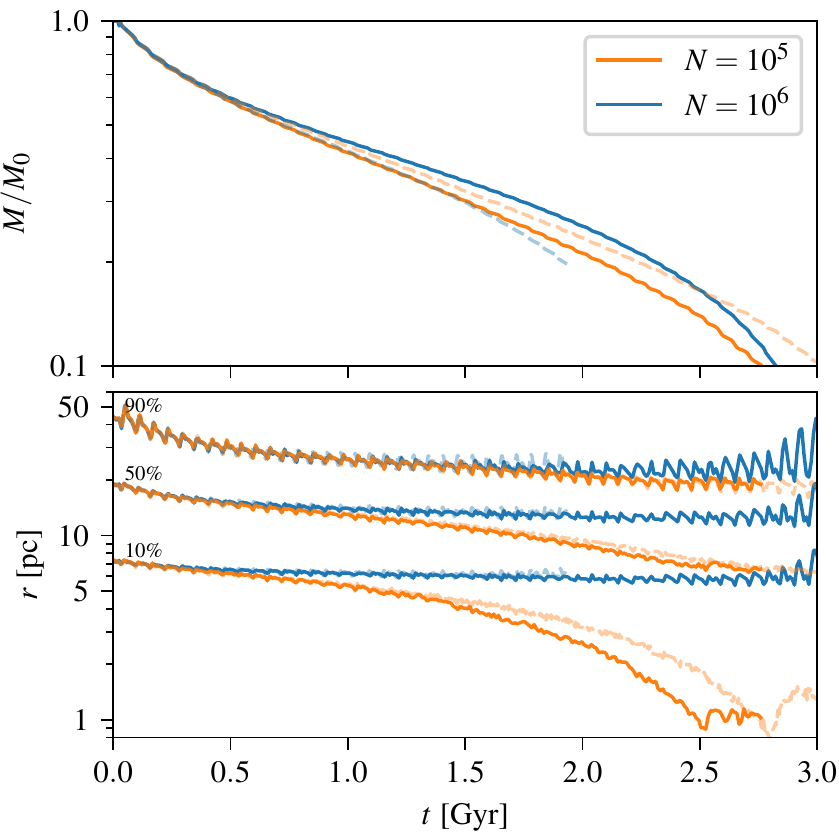}
\caption{Results from the strong tidal field simulation. Upper panel: remaining bound mass for unsoftened (solid lines; performed with \textsc{nbody6++}) and softened models (dashed lines; performed with $\varphi$\textsc{grape}). The blue lines correspond to $N=10^6$ models and orange lines to $N=10^5$ models. Unlike in Figure~\ref{fig:mass-loss}, here we use logarithmic $y$-axis for to accentuate the differences. Lower panel: selected Lagrange radii.}
\label{fig:strong-tide-mass-lagrange}
\end{figure}

In this section we look into the mass evolution of clusters for which the tidal field is dominant (as opposed to relaxation). These models have a $\sim 6$ times smaller semi-major axis than the models discussed in the previous section (the orbital period around the Galaxy is shorter by about the same factor, due to the logarithmic potential). For every point along this orbit, the initial King model's truncation radius is larger than the Jacobi radius (Equation~\ref{eq:tidal-radius}).

The upper panel of Figure~\ref{fig:strong-tide-mass-lagrange} shows the remaining bound mass for the eccentric models in the strong tidal field. All models completely dissolve within less than a Hubble time. The unsoftened models (performed with \textsc{nbody6++}) with $N=10^5$ and $N=10^6$ stars are shown in blue and orange solid lines, respectively (these two curves also appear as the lower curves in Figure~\ref{fig:mass-loss} but shown here in logarithmic scale). The dashed, light-coloured, lines are the softened models (performed with $\varphi$\textsc{grape}). In terms of mass loss, the four eccentric models at this semi-major axis evolve similarly to each other, despite the code difference, presence of softening, and number of particles. The difference in $M(t)$ between the two unsoftened models does not exceed five per cent of the initial mass. As in the weak tide models, the $N=10^6$ model performed with $\varphi$\textsc{grape} shows an unexpected excess in mass loss. This is likely the same effect discussed in Section \ref{sec:bad-model}.

The lower panel of Figure~\ref{fig:strong-tide-mass-lagrange} shows selected Lagrange radii (fewer than in Figure \ref{fig:lagrange-radii-weaktide} in order not to crowd the figure) for these models. Unlike the bound mass evolution, the evolution of the Lagrangian radii however is different, with core collapse occurring in the unsoftened and softened models with $N=10^5$. The softened model's core collapse is shortly delayed as expected; for reference, the core collapse the same model in the weak tidal field occurs at 5~Gyr. The Lagrange radii in the $N=10^6$ cases remain constant until the cluster completely dissolves, indicating that the mass profile evolves self-similarly.

\section{Mass loss fitting}\label{sec:fit}

It is common in the literature to model the mass loss simply by expression of the form
\begin{equation}
\dot{M}\equiv\frac{\mathrm{d}M}{\mathrm{d}t}=-\sum_{i}\frac{M}{\tau_{i}(M/M_{0})^{\gamma_{i}}}\label{eq:mass-loss-diffeq}
\end{equation}
where each term in the sum equals the current mass divided by a characteristic timescale, that in turn is proportional to the mass to some power. The different terms represent physical processes that are assumed to be independent (only coupled through the total mass of the cluster). $\tau_i$ may be effective timescales: \citet{Baumgardt2001} writes a mass loss equation\footnote{\citet{Baumgardt2001} uses $N$ instead of $M$.} for cluster of circular orbits with single term
\begin{equation}
\dot{M}  \propto \frac{M}{t_\mathrm{rlx}^{3/4} t_\mathrm{esc}^{1/4}}
\end{equation}
where $t_\mathrm{rlx}$ is the relaxation time and $t_\mathrm{esc}$ is a characteristic timescale for an unbound star to find its way out of the cluster \citep{Fukushige+2000}, that is thought similar to the crossing time. When factoring in these timescales' mass dependences, \citet{Gieles+2008} get $\gamma=3/4$. \citet{Boutloukos+2003} derive $\gamma = 0.62$ based on observations, while \citet{Lamers+2010} cite a range of $\gamma \sim 0.65$ to 0.85 depending on cluster parameters. \citet{Kruijssen+2011} write a second term associated with tidal shock(s), for which the timescale increases linearly with the half-mass density. Due to the shallow cluster mass--radius relation, this results in a similar mass dependence as the one for two-body relaxation.

When two or more terms are present in the sum in Equation (\ref{eq:mass-loss-diffeq}), no analytical solution exists for every $\gamma_{i}$ and $\tau_{i}$. However if only one term is present, an analytical solution exists. For $\gamma\neq0$,
\begin{equation}
M(t)=M_{0}\left(1-\gamma\frac{t}{\tau}\right)^{1/\gamma}.
\end{equation}
More generally, $M$ at $t=0$ needs not be equal to $M_{0}$. The ``1'' is a constant of integration, any value would still satisfy Equation (\ref{eq:mass-loss-diffeq}). In the special case where $\gamma=0$, where the timescale of the physical process does not depend on the mass, the solution is
\begin{equation}
M(t)=M_{0}e^{-t/\tau}.
\end{equation}

We use Equation (\ref{eq:mass-loss-diffeq}) as a template for fitting our mass loss curves, and do not otherwise delve into the physics behind it or attempt to improve it. Distinguishing different terms using a least-squares fitting procedure to our $M(t)$ curves may not be possible in practice. Our approach is to first attempt fitting with just a single term; if this does not provide a good fit, a new fitting is attempted with two terms. Only in two models, \texttt{33} and \texttt{54}, a two-term fit performed better than a single term.

Another complication is that mass loss in the early stages of our simulations may be due to lack of equilibrium between the stellar system and the tidal field. This effect may not be well described by Equation (\ref{eq:mass-loss-diffeq}), and a non-autonomous correction may be needed to account for it. In the weak tide models, we observe that the mass loss rate starts at zero and initially increases linearly, which is not possible under Equation (\ref{eq:mass-loss-diffeq}), which predicts a non-zero mass loss rate at $t=0$. In the strong tide models, the entire $M(t)$ curve (i.e. including early times) could be fitted with two terms, however this requires large negative values of one of the $\gamma$-s which do not seem to be physically reasonable. \citet{Ernst+2015} suggest a log-logistic curve to describe the dissolution of open clusters in the strong tide case, but it is empirically motivated. We therefore avoid this complication by discarding the first ten Galactic orbits of each model. This choice is arbitrary but seems to give reasonable fits.

\begin{table}
\begin{centering}
\begin{tabular}{|c|c|c|c|c|c|c|}
\hline Model & $\tau_{1}$ & $\gamma_{1}$ & $\tau_{2}$ & $\gamma_{2}$ & $R^{2}$ & $t_{\mathrm{diss}}$\tabularnewline \hline \hline \texttt{1} & $254$ & $2.22$ & \textemdash{} & \textemdash{} & 0.9993 & $115$\tabularnewline \hline \texttt{17} & $42.1$ & $3.03$ & \textemdash{} & \textemdash{} & 09973 & $16.0$\tabularnewline \hline \texttt{25} & $58.1$ & $3.29$ & \textemdash{} & \textemdash{} & 0.9971 & $21.1$\tabularnewline \hline \texttt{33} & $1.84$ & $-0.07$ & $462$ & $3.16$ & 0.9995 & $2.86$\tabularnewline \hline \texttt{49} & $1.57$ & $-0.004$ & $1083$ & $3.06$ & 0.9998 & $2.91$\tabularnewline \hline \texttt{57} & $9.33$ & $1.24$ & \textemdash{} & \textemdash{} & 0.9994 & $5.47$\tabularnewline \hline
\end{tabular}
\par\end{centering}
\caption{Best fitting parameters. The timescales are in Gyr. \label{tab:fits}}
\end{table}

\begin{figure*}
\centering \includegraphics[width=1\textwidth]{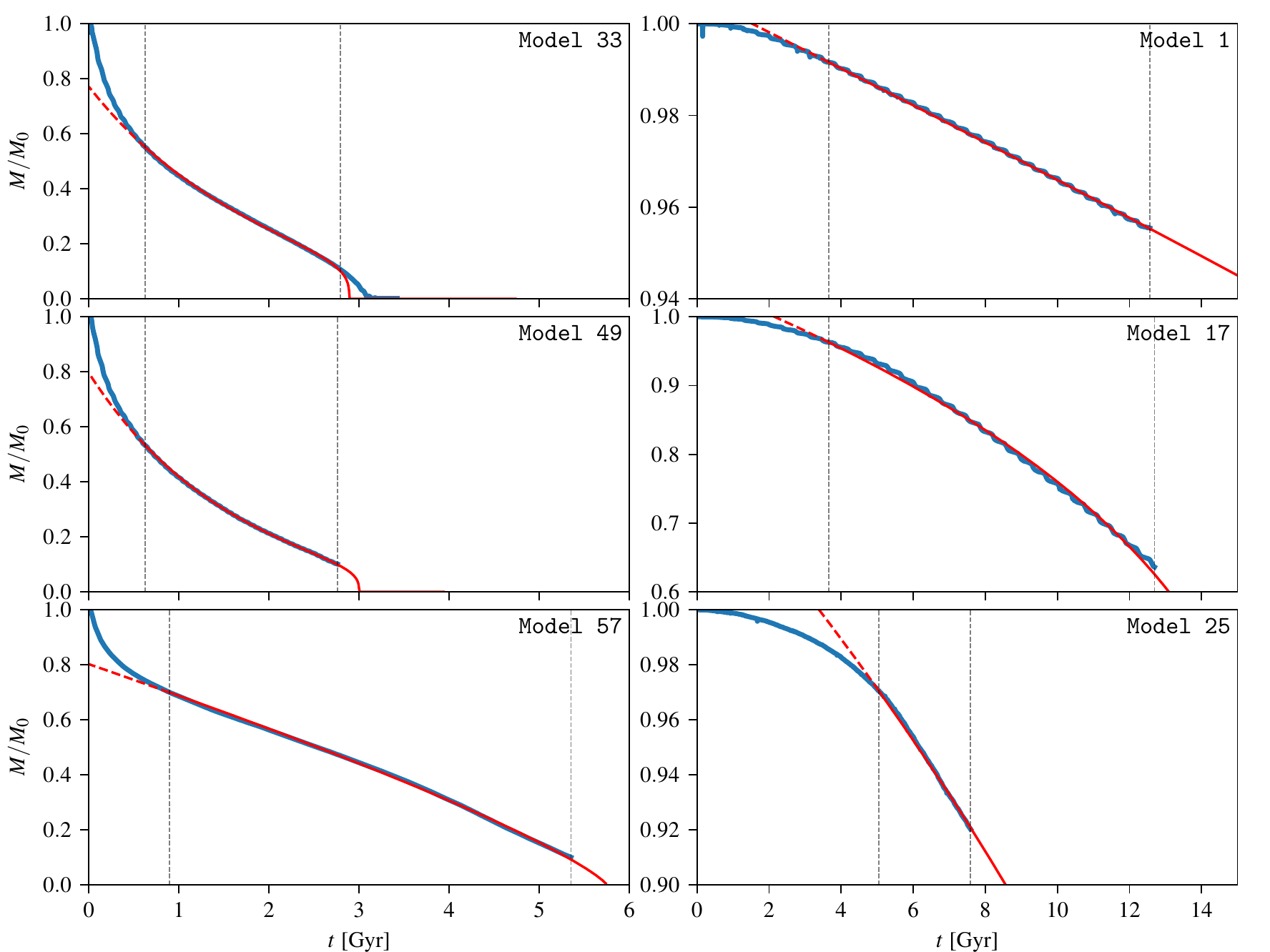}
\caption{Each panel shows the remaining bound mass fraction as a function of time for the six indicated models. The panels on the left correspond to the strong tidal field case (the semi-major axis $a$ is 3.35~kpc) while the panels on the right correspond to the weak tidal field case ($a=19.7$~kpc). The blue lines are the simulation data, while the red lines are the best fitting curves with the parameters shown in Table~\ref{tab:fits}. The vertical grey lines mark the time interval where the fitting was done (starts at 10 Galactic orbits and ends at the end of the simulation data, except for model \texttt{33} which we cut off slightly earlier for numerical reasons).}
\label{fig:fits}
\end{figure*}

Table \ref{tab:fits} shows the best fitting parameters, and Figure~\ref{fig:fits} shows the simulation results in blue and best fitting curves in red (solid lines). The vertical dashed lines show the time interval for the fit, and the dashed red lines are extrapolations of the fitted curves to $t=0$. It is difficult to interpret the powers $\gamma$. In the cases where one term fit was performed, there may very well have been multiple underlying physical processes that could not be distinguished. The four parameters in the solution to the two-term Equation (\ref{eq:mass-loss-diffeq}) are somewhat degenerate (in a practical sense), and we observed that it is sometimes possible to generate similar solutions with quite different timescales and powers. This fact also leads to sensitivity of the minimization procedure to the initial guess. Attempting to force at least one term to have $\gamma=0.7$ however yields a poorer fit than letting the power be a free parameter. While it is prudent not to overinterpret these results due to these numerical difficulties, it is tempting to identify the nearly-zero values of $\gamma_{1}$ for the models in eccentric orbit in a strong tidal field (\texttt{33} and \texttt{54}) with tidal shock.

We define the dissolution time $t_\mathrm{diss}$ as the time it takes for the remaining mass to equal 7 per cent of the initial (post-stellar evolution) total mass. This choice is consistent with the definition in \citet{Baumgardt+2003}. They used 5 per cent, but this was measured against the pre-stellar evolution mass (in their work about 30 per cent of the mass was lost in stellar evolution). The extrapolated values are shown in the right-hand column on Table~\ref{tab:fits}.

From the ratios of the dissolution times of the circular models to the eccentric ones (at the same semi-major axis), we can estimate the constant $c$ from Equation (\ref{eq:Cai}). We get $c=-0.15$ and $1.25$ in the weak and strong tide cases, respectively, as opposed to $c=0.5$ \citep{Cai+2016} and $c=0$ \citep{Baumgardt+2003}. Many differences between our work and the two previously mentioned ones may account for the difference. In particular, the spatial extent of the clusters with respect to the Jacobi radius (e.g. at pericentre) as well as the density model were different and we speculate that these are the most important factors.

\section{Discussion}

We simulated the evolution of massive globular clusters up to 12\,Gyr and explored the effects of particle number, Galactic semi-major axis and eccentricity, and gravitational softening, using three different direct-summation $N$-body codes. We showed that internal and external processes for mass loss can be distinguished according to the rate at which a star's energy changes as it becomes energetically unbound from the cluster. We termed stars ejected due to two-body encounters \textit{kicks} and stars ejected due to the tidal forces \textit{sweeps}, and showed that they can also be distinguished quite well by the radius where they were in the snapshot just prior to their internal energy exceeding zero. Even in the weak tide case, sweeps dominated the mass evolution, but a typical halo globular cluster on a moderately eccentric orbit (such as our model \texttt{1}) has such a low total mass loss rate that it can survive for tens of Hubble times. One must be careful in generalizing this result given the many free parameters in these models, their half-mass radii in particular.

We did not find any difference between the different $N$-body codes at $N=10^5$ despite some large technical differences in their implementation. However we found a numerical artifact at $N=10^6$ which is most likely a result of lack of convergence in the Hermite $\eta$ (time step) parameter, the appropriate value of which may depend on the number of particles. The Ahmad--Cohen neighbour scheme seems to alleviate this problem to some extent.

The main caveat in this work is that we used a spherically symmetric logarithmic potential model which may not be a good approximation to the Galaxy, especially the inner regions. Moreover, our potential is constant over a period of gigayears, again differing from what we understand the Galaxy's evolution to have been.

Yet another caveat is that the potential model we used was completely smooth, while we expect baryonic and dark matter substructure to occasionally interact with globular clusters, potentially heating them. \citet{Webb+2019} however showed that tidal fluctuations from substructure are usually either too small in magnitude or too short in duration to allow stars to energetically escape the cluster. It is not clear however if this conclusion would hold in a scenario where the assembly history of the Galaxy is also taken into account.

In future work, we plan to ``replay'' the cosmological history of a Milky Way-like subhalo from a large cosmological simulation as an external potential to $N$-body simulations of globular clusters. However due to the extremely long dissolution time of our model \texttt{1}, we estimate that typical halo globular cluster on moderately eccentric orbits such may survive with very little change for longer than a Hubble time.

From the analysis perspective, our methodology of distinguishing so-called kicks and sweep based on the escaping stars' energy change rate as they become energetically unbound, seems to be robust and may be applicable more broadly in stellar dynamics. For example, in galactic centre simulations, where escapes may occur due to interactions with intermediate-mass black holes. This analysis reveals that even in the weak tidal field case, two-body encounters are directly responsible for only a very small fraction of escaping stars (i.e. the kicks). Two-body relaxation as a process, however, is very important for the sweep and total mass loss rate as well, as it determines how quickly the reservoir of loosely bound stars is refilled.

Our models \texttt{1} and \texttt{17} are models of a globular cluster on a moderately eccentric orbit in the weak tidal field case. They are identical but for the number of stars, $N=10^6$ and $10^5$, respectively. While the details of their mass evolution are different, Chandrasekhar's theory does provide a good approximation to the ratios of kicks and sweeps between the two models and their fitted dissolution times. Our models \texttt{33} and \texttt{49} are similarly differing models but in the strong tidal field case, where the tidal interaction with the Galaxy is so strong that two-body relaxation plays almost no role in their mass evolution, and their dissolution times are nearly identical.

\section*{Data availability}
The data underlying this article will be shared on reasonable request to the corresponding author.

\acknowledgments

This work was supported by the Deutsche Forschungsgemeinschaft (DFG, German Research Foundation) -- Project-ID 138713538 -- SFB 881 (`The Milky Way System'), by the Volkswagen Foundation under the Trilateral Partnerships grants No. 90411 and 97778.

PB and RS acknowledges support by the Chinese Academy of Sciences through the Silk Road Project at NAOC, the President's International Fellowship (PIFI) for Visiting Scientists program of CAS, the National Science Foundation of China under grant No. 11673032.

The work of PB was supported under the special program of the NRF of Ukraine ``Leading and Young Scientists Research Support'' - ``Astrophysical Relativistic Galactic Objects (ARGO): life cycle of active nucleus'',  No. 2020.02/0346.

\bibliographystyle{yahapj}
\bibliography{main}

\end{document}